\documentclass[aps,prb,twocolumn]{revtex4}

\usepackage{graphicx}
\usepackage{dcolumn}
\usepackage{amsmath}
\usepackage{color}
\usepackage{soul}
\usepackage{gensymb}

\newcommand{\TTF}{TTF[Ni(dmit)$_2$]$_2$ }

\begin{document}

\title{Amplitude mode of charge density wave in \TTF observed by electronic Raman scattering}

\author{M. Revelli Beaumont$^{1,2}$}
\author{Y. Gallais$^1$}
\author{A. Sacuto$^1$}
\author{K. Jacob$^{2}$}
\author{L. Valade$^{2}$}
\author{D. de Caro$^{2}$}
\author{C. Faulmann$^2$} 
\author{M. Cazayous$^{1}$}\thanks{corresponding author : maximilien.cazayous@u-paris.fr}
\affiliation{$^1$Laboratoire Mat\'eriaux et Ph\'enom\`enes Quantiques (UMR 7162 CNRS), 
	Universit\'e de Paris, 75205 Paris Cedex 13, France\linebreak
	$^2$Laboratoire de Chimie de Coordination (UPR 8241), Universit\'e Paul Sabatier Toulouse, France}
	
\begin{abstract}
We measured the optical signature of the charge density waves (CDWs) in the multiband conductor \TTF by electronic Raman scattering. At low energies, a hump develops below 60 K. This hump is associated to the amplitude mode of the CDW with an energy around 9 meV. Raman symmetry-resolved measurements show that the CDW amplitude mode is anisotropic and that the CDW can be associated to the band nesting of Ni(dmit)$_2$ chains. 
\end{abstract}

\maketitle

Charge density waves (CDWs) are a widespread physical phenomenon in condensed matter and are observed in many solids, especially in low-dimensional systems.\cite{Gruner} CDWs can be grouped into three main categories.\cite{Zhu, Martin}

The first category corresponds to CDWs that have their origin in the instability that occurs below a T$_{CDW}$ critical temperature, as described by Peierls.\cite{Peierls} A Peierls instability is a dimerization of the lattice crystal and the opening of an energy gap at the Fermi level associated to an electronic transition from metal to insulator. The decrease of the electron energy modulates the electron density of the system. It generates the spontaneous modulation of the crystal lattice by the electron-phonon interactions.\cite{Kohn, Mazin} In the Peierls picture, the Fermi surface nesting with a vector q$_{CDW}$ is at the origin of CDWs resulting in a strong peak in the susceptibility at q$_{CDW}$ and a sharp dip, the so-called Kohn anomaly, in phonon dispersion at 2q$_{CDW}$. 

In the second group, CDWs originate from electron-phonon interactions but are not driven by the nesting and the transition is not associated to a metal-insulator transition. The signature of the transition results in a phonon mode at q$_{CDW}$ with an energy going to zero at T$_{CDW}$. Among dichalcogenides, NbSe$_2$ is an example of this category. 

The third categorie describes CDWs associated to a charge modulation without evidences for nesting or electron-phonon interactions. Such a behavior for the CDWs is observed in unconventional superconductors such as cuprates.\cite{Chang,Comin,Loret,Lee}

A CDW state is described by an amplitude mode (amplitudon) and a phase mode (phason).\cite{Lee} The phason corresponds to the vibration of the electron density wave in a rearranged lattice and the amplitudon modulates the magnitude of the gap leading to oscillations of the amplitude order parameter. The oscillation frequency corresponds to the phonon mode driving the nesting.
For several years now, CDWs have emerged as a phenomenon that interacts or competes with other orders (superconductivity, ...) which has led to a strong renewal of interest for this wave.\cite{Monceau} It is still in NbSe$_2$ that one find a good example of the interaction between orders. In this compound, superconductivity coexists with a CDW order. Spectroscopic probes point out the so-called "Higgs" mode which becomes active by removing spectral weight from the CDW amplitude mode upon entering the superconducting state.\cite{Grasset,Tsang} The coupling between the CDW and the Higgs mode is made possible by the fact that the amplitude mode of the CDW "shakes" the density of states at the Fermi level and then modulates the amplitude of the superconducting order parameter.\cite{Cea}

\begin{figure}[h!]
	\begin{center}
		\includegraphics[width=7.5cm]{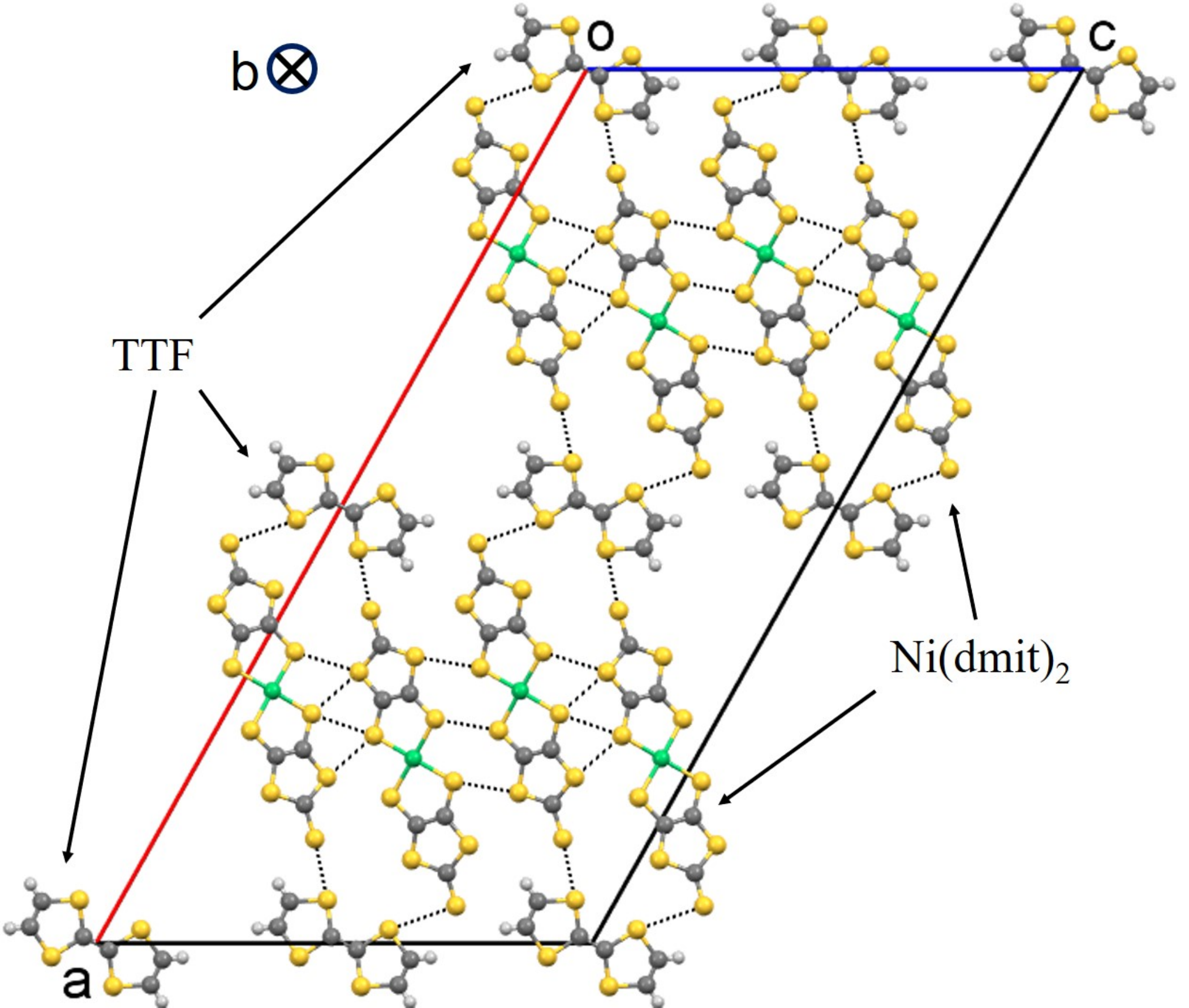}
		\caption{Projection onto the (010) plane of the \TTF crystal structure. Slabs of Ni(dmit)$_2$ alternate with slabs of TTF.}
		\label{Fig1}
	\end{center}
\end{figure}

When one thinks to one-dimensional systems and superconductivity, the physics of organic superconductors quickly comes to mind. 
The superconductivity\cite{Parkin} and CDW\cite{Ravy} has been observed in (BEDT-TTF)$_2$ReO$_4$. It is in this compound that the competition between the two orders has been first suspected.\cite{Kaddour}
 A phenomenom that fosters competition between orders is the contribution to the Fermi level of molecular orbitals. Such a situation is encountered in the TTF[M(dmit)$_2$]$_2$ family in which TTF is the tetrathiafulvalene molecule and dmit$_2$ is the 2-thioxo-1,3-dithiole-4,5-dithiolato anion (M is the Ni, Pd or Pt atom).\cite{Canadell} The present work focuses on \TTF. A comparison between several one-dimensional compounds with a CDW such as TTF-TCNQ, BaVS$_3$ or blue bronze K$_{0.3}$MoO$_3$ can be found in the review article of J-P. Pouget and in the book of D. J\'erome and L. G. Caron.\cite{Pouget, Garon}

Figure \ref{Fig1} shows a projection on the ac plane of the \TTF structure. In this plane, TTF molecules alternate with blocks of Ni(dmit)$_2$ molecules along the crystallographic a axis. These molecules stack up in the crystallographic direction b to give rise to columns of TTF cations alternating with columns of Ni(dmit)$_2$ anions. The structure can be then described as alternating layers of TTF and Ni(dmit)$_2$. The b axis is the crystal elongation axis.\cite{Cassoux} Brossard {\it et al}. discovered that this material becomes superconducting at
T$_C$ = 1.6 K under a pressure of 7 kbar.\cite{Brossard} At ambient pressure, 1D CDW 
fluctuations along the stacking direction b with the wave vector q = 0.40(2)b$^*$ has been observed by X-ray diffuse scattering experiments and the
existence of possible successive CDW transitions below 60 K has been proposed.\cite{Ravy89} In this phase, the metallic state coexists with a CDW state with its own nesting wave vector. More recently, two successive CDW transitions have been identified around 55 K and 35 K and associated to the Ni(dmit)$_2$ chains comparing resistivity measurements under pressure and band structure calculations.\cite{Kaddour}

In this letter, we are able to directly detect by Raman spectroscopy the CDW amplitude mode and to determine its appearance temperature and its energy. Furthermore, polarization analyses show that the CDW is associated to the Ni(dmit)$_2$ chains.

\begin{figure}[h!]
	\begin{center}
		\includegraphics[width=8.5cm]{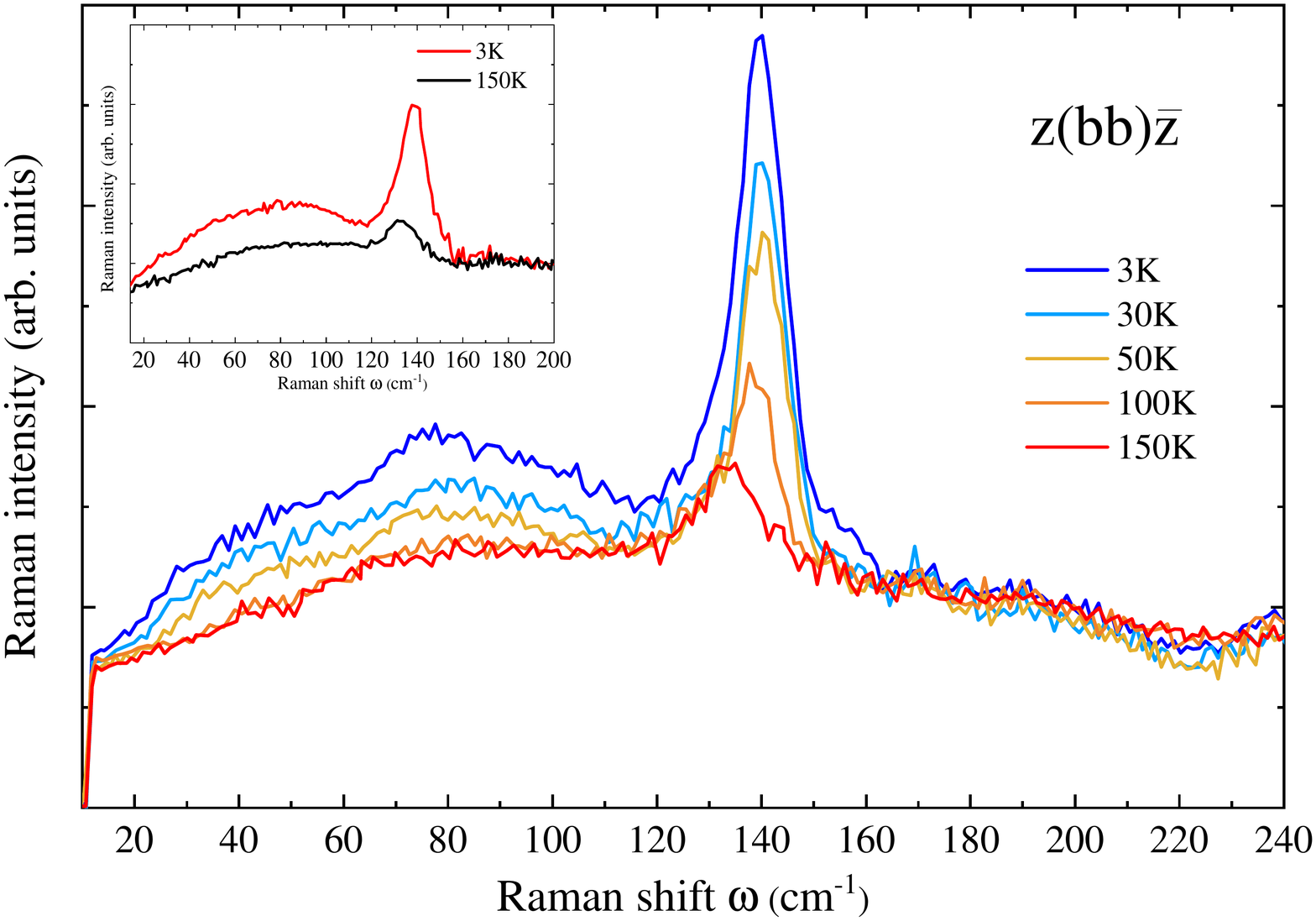}
		\caption{Raman spectra of \TTF measured between 10 and 240 cm$^{-1}$ with the z(bb)\={z} configuration as a function of temperature. Only a few spectra have been plotted for an easier visualization of changes. The inset shows the Raman spectra measured at 3 and 100 K using the laser wavelength $\lambda$ = 660 nm and the same polarizations.}
		\label{Fig2}
	\end{center}
\end{figure}

Single crystals are obtained by a slow interdiffusion of saturated solutions of (TTF)$_3$(BF$_4$)$_2$ and (\textit{n}-Bu$_4$N)[Ni(dmit)$_2$].\cite{Bousseau} The crystal has the monoclinic symmetry and crystallizes in the C 2/c space group. High quality \TTF single crystals have a typical size 2 $\times$ 0.15 $\times$ 0.075 mm$^3$ with the largest dimension along
the b direction. Raman scattering measurements have been performed using a triple spectrometer Jobin Yvon
T64000 equipped with a liquid-nitrogen-cooled CCD detector. 
The incident laser spot is about 50 $\mu$m diameter size and the power is equal to 2 mW, small enough to keep the laser heating as low as possible and large enough to improve the ratio of the signal from the sample to the noise. The laser lines used to probe the sample, from solid state lasers, are at 532 nm and 660 nm. Measurements between 3 and 300 K have been performed using an Cryomech closed-cycle He cryostat.
The Raman spectra have been measured in backscattering using two optical configurations z(bb)\={z} and z(cc)\={z}.\cite{Porto1966} The incident wave vector is anti-parallel to the scattered one and both are along z axis. The incident and scattered polarizations of the light are along the b and c axis of the sample.

\begin{figure}[h!]
	\begin{center}
		\includegraphics[width=7.5cm]{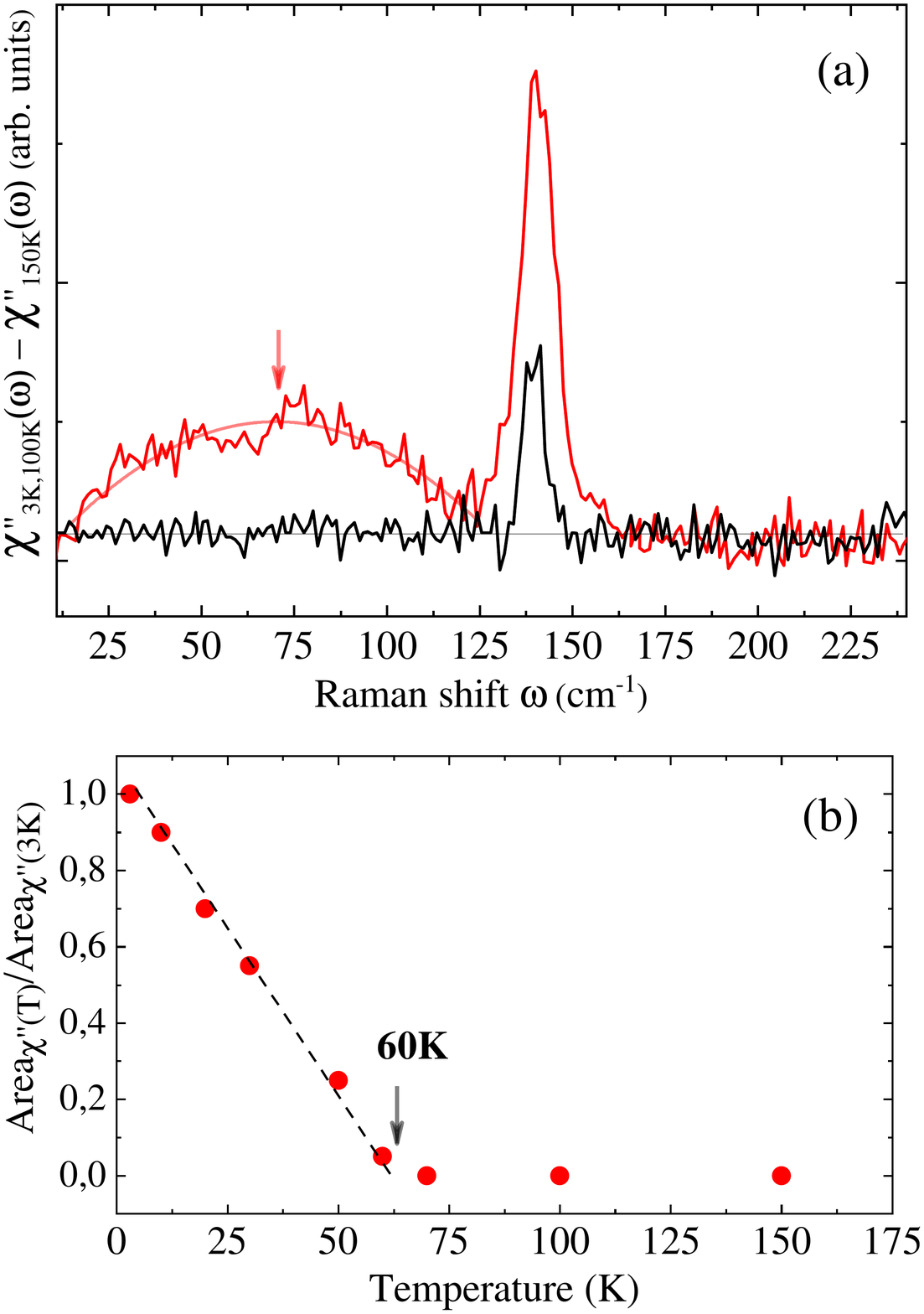}
		\caption{(a) Raman spectra at 3 K and 100 K substrated from the one at 150 K, (b) Normalized area of the hump at 70 cm$^{-1}$ as a function of temperature.}
		\label{Fig3}
	\end{center}
\end{figure}

Figure \ref{Fig2} shows the Raman spectra measured at low energies between 5 K and 150 K using the z(bb)\={z} configuration. 
At high energy beyond 160 cm$^{-1}$ the spectra are superimposed without additional modifications (vertical translation, ...). At 3 K, we observe a phonon mode at 140 cm$^{-1}$ and a lower energy hump centered around 70 cm$^{-1}$. The phonon modes of \TTF can be found in Ref. \onlinecite{Valade}. 
In our measurements, we do not observe phonons below 100 cm$^{-1}$ whereas intermolecular modes are expected. Notice that there are no studies or calculations showing phonons on TTF[Ni(dmit)$_2$]$_2$ below 100 cm$^{-1}$. Their absences have to be noticed and might be explained by their intensities which are too low to be detected or by resonance effects.
The phonon signal is might be also superimposed on the signal of the charge density wave. This last point might explain the shape of the hump associated with the CDW around 80 cm$^{-1}$. However, they should appear above the temperature of the CDW in Fig. \ref{Fig2} and with the subtraction of the spectra in Fig. \ref{Fig3} which is not the case. 
The phonon mode at 140 cm$^{-1}$ corresponds to the whole Ni(dmit)$_2$ deformation mode.\cite{Pokhodnya} As the temperature increases, the phonon frequency decreases as well as its intensity due to the thermal expansion of the lattice. In the same temperature range, the estimated softening with temperature of the hump  is about 5 cm$^{-1}$ but corresponds to the uncertainty due to the width of the hump. Its intensity decreases until the spectra overlap above 100 K.

The Raman response at 3 K ($\chi^{"}_{3K}$) and at 100 K ($\chi^{"}_{100K}$) subtracted from the one at 150 K ($\chi^{"}_{150K}$) have been plotted in Fig. \ref{Fig3}(a). The $\chi^{"}$($\omega$) response is obtained by correcting all Raman spectra for the spectrometer response and the Bose factor. The Bose factor linking the Raman intensity and the imaginary part of the Raman response function stems from the fluctuation dissipation theorem and is a general property of inelastic scattering techniques.\cite{Hayes} The electronic Raman response is proportional to the imaginary part of the density-density correlation function of the material. In order to access to this quantity, we need to divide the experimental Raman response from the Bose factor and made the correction of spectrometer response.

This difference clearly highlights the low energy hump and allows us to determine its energy a slightly above 70 cm$^{-1}$. The analysis of the area under this hump as a function of temperature shows that this excitation disappears around 60 K $\pm$ 5 K as shown in Fig. \ref{Fig3}(b).

Notice that X-ray measurements reported by S. Ravy {\it et al.} show satellite reflections below 60 K pointing out the appearance of a CDW in \TTF.\cite{Ravy89} The CDW transition has been recently identified around 55 K.\cite{Kaddour} Our temperature measurements indicate that the hump observed at low energies in the Raman spectra corresponds to the signature of the CDW mode.
However, this hump might be associated at first glance to another origin as a phonon mode (inter-molecular phonon mode or a finite q-phonon activated by the CDW) or luminescence for example. The width of the phonons is generally in the order of a few cm$^{-1}$ at low temperatures in single crystals as shown in Fig. \ref{Fig2} with the full width at half maximum (FWHM) of the 140 cm$^{-1}$ phonon equals to 10 cm$^{-1}$. The width of the hump (FWHM around 60 cm$^{-1}$) does not plead for a phonon, in contrast to the width of amplitude mode in 2H-TaS$_2$ for example measured around 40 cm$^{-1}$.\cite{Grasset} To rule out a luminescence signal that would appear unfortunately at 60 K, we have performed measurements with an another laser wavelength $\lambda$ = 660 nm (inset of Fig. \ref{Fig2}). The hump doesn’t shift in energy with the laser wavelength allowing to exclude this possibility as an origin of the hump.

The question is now to know whether this hump is associated to the gap or to the amplitude mode of the CDW. The amplitude mode corresponds to the vibration of the ions due to the intensity oscillation of the maximum/minimum charge density and the magnitude of the CDW gap.\cite{Lee, Rice}
A first argument is given by the expected signature of a gap in Raman spectra. A gap opening is characterized by a transfer of spectral weight in the electronic background from low to high energy. Figures \ref{Fig2} and \ref{Fig3}(a) clearly show that this is not the case : the hump is decreasing in intensity  with temperature without exhibiting a low energy spectral weight recovery.
A second argument can be obtained by comparing our measurements to mean field theory.
The Raman peak associated to the gap $\Delta_{CDW}$ of the CDW is expected at 2$\Delta_{CDW}$.\cite{Vanyolos} If we associate the hump of Fig. \ref{Fig2} with the gap of the CDW, it means that $\Delta_{CDW}$ is about 70 cm$^{-1}$, equivalent to 9 meV. 
As a first approximation, we can use mean field theory to relate the gap 
$\Delta_{CDW}$ to T$_{CDW}$ {\it i.e.} 2$\Delta_{CDW}$ = $\alpha k_B T_{CDW}$.
In 1D system, $\alpha$ has an expected value much higher than the theoretical value equal to 3.52. The experimental ratio gives 1.7, twice as low as the theoretical ratio. The observed hump is then unlikely related to the gap of the CDW. One possibility remains, the observed hump in our Raman spectra could be associated to the amplitude mode of the CDW order that appears below T$_{CDW}$ = 60 K. The fact that the amplitude mode softens little with temperature can be observed in transition metal dichalcogenide for example.\cite{Grasset} Deviation from the expected mean-field behavior for the order parameter is not surprising in low dimensional systems. As observed in other systems,\cite{Tsang,Measson} Raman spectroscopy is sensitive to the amplitude mode of the charge wave and not directly to the gap.

\begin{figure}[h!]
	\begin{center}
		\includegraphics[width=7.5cm]{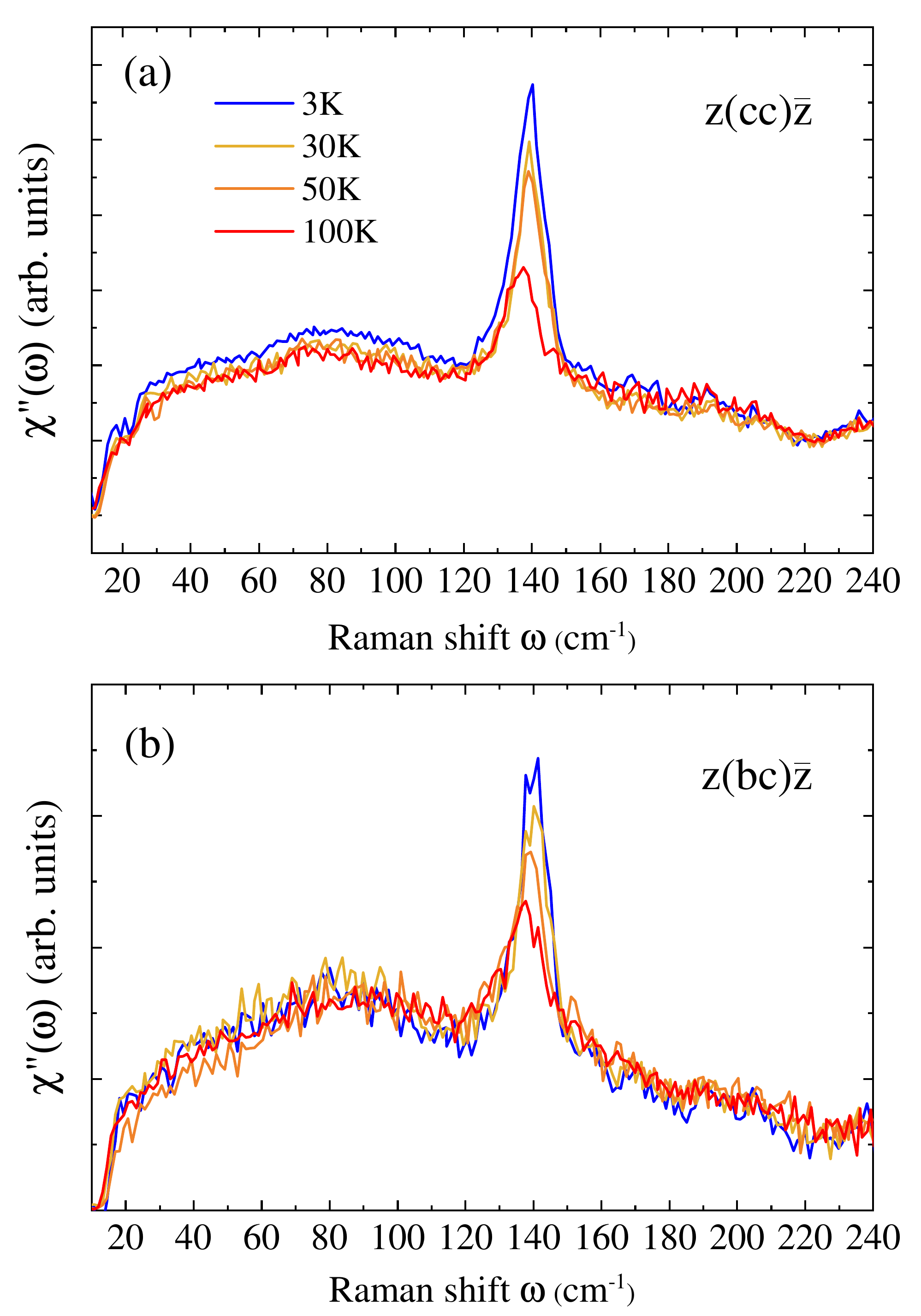}
		\caption{Raman spectra with (a) incident and scattered polarizations along the c axis (z(cc)\={z} configuration) and (b) crossed polarizations configuration (z(bc)\={z}) as a function of temperature.}
		\label{Fig4}
	\end{center}
\end{figure}

In addition, the CDW has been measured with the z(cc)\={z} and z(bc)\={z} configurations in Fig. \ref{Fig4}(a) and (b) respectively. Since the intensity of the hump is very weak in Fig. \ref{Fig4}(a) and drops rapidly with increasing temperature, we are not able experimentally to show that the hump exists between 30 and 60 K. The spectra are almost superimposable beyond 30 K. 
The hump does not exist in cross polarizations. These measurements show that the Raman response of the CDW amplitude mode is anisotropic and almost exclusively along the b axis. Calculated electronic Raman cross section in quasi-one-dimensional interacting-electron systems with density wave ground state have shown that collective contributions to the Raman response appear only with polarizations along the chains associated to the CDW.\cite{Vanyolos} The b axis corresponds here to the directions of the Ni(dmit)$_2$ stacks.

The observation of the CDW signature perpendicular to the Ni(dmit)$_2$ stacks in Fig. \ref{Fig4}(a) is not expected due to the one-dimensionality of the CDW in this compound. This signal might be a remnant of the signal along the b axis due to the fact that we are not perfectly along the c axis.  We also note that if the CDW is along the Ni(dmit)$_2$ stacks (b axis), it is not obvious that the Raman signature of the CDW appears only with polarizations along the b axis. For example, the amplitude mode in NbSe$_2$ is observed using two different polarization configurations (A$_1$ and E).\cite{Sooryakumar} Naively it is expected to appear with only one polarization configuration, the fully symmetric A$_1$. The reason for the disagreement is not yet well understood, but an alternative explanation has been proposed based on anharmonic effects.\cite{Klein}

TTF[Ni(dmit)$_2$]$_2$ is a quasi-1D material with uncorrelated spins on the Ni(dmit)$_2$ stacks.
Bourbonnais $_2$ performed nuclear magnetic resonance (NMR) measurements on TTF stacks and shown that the TTF chains keep a metallic behavior down to low temperatures without gap opening.\cite{Bourbonnais} So the CDW cannot be associated to these chains. On the other hand, NMR measurements on Ni(dmit)$_2$ chains attribute the CDW to these chains.\cite{Vainrub}
The CDW correlation length along the chains of the order of b at room temperature increases above 20 nm at 55 K. However, there is no interchain CDW correlations up to 55K. Below this temperature the authors show that TTF[Ni(dmit)$_2$]$_2$ undergoes at 55 K a CDW transition related to the nesting of the LUMO band of the Ni(dmit)$_2$ stacks. Notice that the CDW lateral order is not perfect probably due to the Coulomb coupling between CDWs.
The authors show that TTF[Ni(dmit)$_2$]$_2$ undergoes at 55 K a CDW transition related to the nesting of the LUMO band of the Ni(dmit)$_2$ stacks and a second CDW transition at 35 K associated to the nesting of the HOMO bands. The insulating ground state is thus considered to be CDW of the Ni(dmit)$_2$ stacks with different wave vectors. Consequently, our measurements support the idea that CDW is associated to the Ni(dmit)$_2$ stacks.

What are the prospects of the direct measurement of CDW in an organic conductor ? Raman experiments contributed significantly to the establishment of the d-wave nature of the order parameter in high-temperature superconductors.\cite{Devereaux} This technique also allows to investigate the temperature and pressure dependence of the CDW amplitude mode in several systems.\cite{Wilson, Friend} In M(dmit)$_2$-based compounds, superconductivity develops under pressure in competition with a CDW ground state.\cite{Brossard, Brossard89} The possibility to probe superconductivity and the CDWs as shown in this work with the same technique will allow to study more efficiently the interaction between the two orders in organic compounds.

To conclude, we have optically highlighted the CDW signature in the molecular conductor \TTF. This allowed us to determine precisely the energy of the CDW amplitude mode and its temperature of appearance. The polarization measurements show that the
CDW amplitude mode is very anisotropic and that the CDW originates from the Ni(dmit)$_2$ stacks.



\end{document}